# Multimodal Insights into Credit Risk Modelling: Integrating Climate and Text Data for Default Prediction

Zongxiao Wu[a], Ran Liu[b], Jiang Dai[a, *], Dan Luo[c]

[a] Business School, University of Edinburgh, Edinburgh, EH8 9JS, United Kingdom

[b] Edinburgh Business School, Heriot-Watt University, Edinburgh, EH14 4AS, United Kingdom

[c] Henley Business School, University of Reading, Reading, RG6 6UD, United Kingdom

**Abstract**

Credit risk assessment increasingly relies on diverse sources of information beyond traditional structured financial data, particularly for micro and small enterprises (mSEs) with limited financial histories. This study proposes a multimodal framework that integrates structured credit variables, climate panel data, and unstructured textual narratives within a unified learning architecture. Specifically, we use long short-term memory (LSTM), the gated recurrent unit (GRU), and transformer models to analyse the interplay between these data modalities. The empirical results demonstrate that unimodal models based on climate or text data outperform those relying solely on structured data, while the integration of multiple data modalities yields significant improvements in credit default prediction. Using SHAP-based explainability methods, we find that physical climate risks play an important role in default prediction, with water-logging by rain emerging as the most influential factor. Overall, this study demonstrates the potential of multimodal approaches in AI-enabled decision-making, which provides robust tools for credit risk assessment while contributing to the broader integration of environmental and textual insights into predictive analytics.

[*] Corresponding author.
Email addresses: Zongxiao.Wu@ed.ac.uk, Ran.Liu@hw.ac.uk, Jiang.Dai@ed.ac.uk, dan.luo@henley.ac.uk

# 1. Introduction

Credit risk refers to the risk that a borrower will default on a debt by failing to make the required repayments (Fu et al., 2021). With the ability to depict the creditworthiness of borrowers, credit scoring helps improve returns and financial stability and thus is undoubtedly of major concern for both financial institutions and individual investors. From an information systems (IS) perspective, credit scoring is a core decision support task, where system effectiveness depends on the quality, diversity, and interpretability of the information inputs used to inform human and algorithmic decisions (Phillips-Wren, 2013; Watson, 2017; Wang et al., 2023). However, due to the inevitable information asymmetry between lenders and borrowers, credit risk evaluation has always posed a considerable challenge, particularly for micro and small enterprises (mSEs), whose financial records are often incomplete, infrequent, or unreliable (Wu et al., 2025).

Recent advances in data availability and computing technologies have expanded the scope of artificial intelligence (AI)-based decision support systems for credit risk assessment (Wang et al., 2020; Li et al., 2024). Beyond traditional structured credit variables, lenders can now access a wide range of alternative data sources, including text, audio, image, video, satellite observations, and so forth. These developments have created new opportunities for multimodal credit scoring, where heterogeneous data modalities can be jointly modelled to construct richer representations of credit risk. At the same time, many risk-relevant signals, particularly those related to environmental exposure and borrower behaviour, remain insufficiently integrated into existing scoring workflows, which limits the transparency and effectiveness of credit decisions (Vössing et al., 2022).

Among alternative data sources, climate data has gained growing attention due to its significant influence on borrowers' operating conditions and repayment capacity, especially in climate-sensitive sectors such as agriculture, real estate, and logistics (Calabrese et al., 2024; Lane, 2024). Climate shocks, including droughts, floods, and temperature extremes, can disrupt production, damage assets, increase costs, and weaken cash flows, thereby increasing default risk (Addoum et al., 2023; Aguilar-Gomez et al., 2024). These effects are particularly pronounced for mSEs, which often operate with thin margins and limited buffers. Despite this relevance, most existing studies examine climate risks in isolation, typically focusing on single events or indicators and relying on conventional statistical models. This limits their ability to capture interactions across multiple climate conditions and other borrower-specific risk factors (Castro & Garcia, 2014; Kaur Brar et al., 2021). In parallel, textual data has emerged as an important source of soft information in credit scoring. Using natural language processing (NLP) techniques, prior research shows that narratives, such as loan officer assessments or borrower descriptions, contain valuable signals related to borrower repayment intension, behaviour, and local conditions that are difficult to quantify using structured data alone (Stevenson et al., 2021; Kriebel & Stitz, 2022; Wu et al., 2025). From a socio-technical systems perspective, such text data represents human judgement embedded in organisational processes and thus can provide complementary information that enhances machine-supported decision-making.

However, while structured credit data, climate data and text data have each demonstrated predictive value when modelled separately, their joint potential remains largely unexplored. These modalities differ significantly in structure, dimensionality, temporal resolution, and noise characteristics. Naïve fusion strategies, such as simple feature concatenation, may underutilise temporal climate variation, overweight high-dimensional text embeddings, or fail to capture cross-modal interactions. As a result, such approaches often yield limited performance gains and poor interpretability (Korangi et al., 2023; Tavakoli et al., 2025). Multimodal learning frameworks, by contrast, provide a possible solution by encoding each modality through dedicated representation learning components and combining them through fusion mechanisms. This design enables the model to down-weight noisy or weakly informative inputs, exploit complementarities across data sources, and learn interaction effects that are not obtained from any single modality alone.

Motivated by these considerations, we propose a multimodal learning framework that employs modality-specific encoders and adopts a representation-level intermediate fusion mechanism to combine heterogeneous representations for credit default classification (Tavakoli et al., 2025). We implement this framework using state-of-the-art deep learning models, including long short-term memory (LSTM), gated recurrent units (GRU), and transformer models, which are well suited to modelling temporal dependencies and complex semantic patterns. Our empirical analysis is based on a unique dataset of 4,172 agricultural mSE loans, which includes structured borrower characteristics, loan officer textual assessments, and four climate risk factors measured over the 12 months preceding loan origination. Through comparisons among these three types of data, our unimodal results first demonstrate that models relying on either climate or text data outperform the structured-only model, with climate-only models showing particularly strong performance in this agricultural setting. When considering multimodal models, the results show that integrating different data types significantly outperforms unimodal models, which highlights the advantages of information fusion in terms of enhancing credit default prediction. Although deep learning models have been demonstrated to improve the accuracy of model predictions, interpreting how these predictions are derived remains a significant challenge. The correlation analysis of model outputs highlights the critical role of climate data in multimodal predictions. Using Shapley additive explanations (SHAP), we further identify the relative contributions of climate risk factors and find that water-logging by rain risk is the most influential predictor, followed by other factors, each of which interacts uniquely with borrower characteristics.

Our study makes three main contributions to the field. First, we advance IS research on AI-enabled decision support for credit scoring by proposing a state-of-the-art multimodal framework that integrates heterogeneous data sources (Watson, 2017). This framework is distinguished by its flexibility and can be extended to other data modalities and settings. Our framework also provides transparent, visualisable evidence on the incremental contribution of each information source, thereby enhancing the auditability and practical usability of model outputs for effective human-AI collaborations (Abedin et al., 2022; Vössing et al., 2022). Second, our study complements and contributes to the existing literature on the value of alternative data for predicting important outcomes in financial markets. The findings demonstrate that including such multimodal data can significantly improve the accuracy of loan default prediction for mSEs, whose structured data (i.e., financial information) is often unavailable to lenders. Third, this study enriches the literature on the integration of climate data into business decision-making, which demonstrates how multiple climate risk factors jointly influence borrower repayment behaviour (Castro & Garcia, 2014; Pelka et al., 2015; Römer & Musshoff, 2018). This thereby highlights the importance of incorporating environmental data into operational credit decision systems.

The remainder of the paper is organised as follows. In Section 2, we review the related literature on the integration of climate data, textual information, and multimodal learning approaches in credit risk management, and identifies the key research gaps addressed in this study. Section 3 provides a detailed description of our dataset. Section 4 outlines the proposed multimodal architecture and the benchmark models used in this architecture. Section 5 discusses our experimental design, performance metrics, and interpretability methods. Section 6 presents the empirical results and explores which climate risk factor contributes most to the prediction performance. Finally, Section 7 summarises the contributions and suggests ideas for future work.

## 2. Literature Review

In this section, we organise the existing literature into three main strands and outline the research gaps that motivate our work. Section 2.1 summarises prior studies on integrating climate data into credit risk management. Section 2.2 reviews recent studies on the use of text data in credit default prediction. Section 2.3 discusses recent developments in multimodal credit risk modelling.

*2.1 Integration of climate data in credit risk management*

Credit scoring refers to an automatic credit assessment process that helps lenders identify borrowers who will fail to fulfil their financial obligations within a group of loan applicants. Given the surge in the consumer lending market, novel credit scoring models continue to attract considerable interest in academia and industry (Song et al., 2023; Wu et al., 2025). In recent years, the growing impacts of global climate change have drawn increasing attention to understanding how climate risks affect credit risk management.

Conceptually, climate risk can affect creditworthiness through several economic channels. First, climate shocks directly impair firms' operational capacity. Extreme temperatures, droughts, and excessive rainfall can disrupt production schedules, reduce labour availability, damage equipment or inventory, and delay logistics. These disruptions reduce output and revenue while increasing repair and input costs, thereby tightening liquidity and increasing the likelihood of short-term cash-flow shortages (Addoum et al., 2023; Aguilar-Gomez et al., 2024). For mSEs that typically operate with thin margins and limited working capital, such shocks can have disproportionately severe effects. Second, climate shocks can trigger broader market-level disruptions that weaken demand and strain supply chains. Extreme weather events may depress local consumption, interrupt service delivery, or temporarily shut down upstream and downstream partners. Small firms, particularly those dependent on local networks and lacking diversified customer bases, are especially vulnerable to these disruptions. Third, adverse climate events often coincide with tighter credit conditions. Severe climate events can lead financial institutions to reallocate credit, tighten lending standards, or raise collateral requirements in affected regions (Calomiris et al., 2017; Gutierrez et al., 2023). These reductions in credit supply make it more difficult for firms, especially mSEs, to obtain external financing. Taken together, these mechanisms show that climate risks can undermine firm performance by reducing cash flows, increasing their volatility, constraining access to credit, and accelerating the deterioration of assets. As a result, mSEs with higher exposure to unfavourable climate conditions are more likely to face heightened credit risk and a higher probability of default (Aguilar-Gomez et al., 2024; Duong et al., 2025).

Consistent with these mechanisms, empirical studies have shown much evidence that climate events can place borrowers in disadvantaged positions, which include weakening credit quality (Collier et al., 2011), reducing collateral values (Nguyen et al., 2022), raising borrowing cost (Delis et al., 2024), limited access to credit (Berg & Schrader, 2012), and increasing the risk of default (Calabrese et al., 2024; Lane, 2024). The severity of these impacts varies across industries and is particularly pronounced in climate-vulnerable sectors such as agriculture. For example, Ouazad (2022) shows that extreme weather reduces agricultural productivity, thereby diminishing borrowers' repayment capacity and increasing non-performing loans. Aguilar-Gomez et al. (2024) demonstrate that extreme temperatures elevate default rates, with agricultural borrowers most affected. de Roux (2021) reports that there is a significant positive relationship between extreme precipitation and loan default among Colombian coffee farmers. Similarly, climate is also a critical input to the logistics sector, where transport firms may experience declining sales, reduced profitability, and rising credit risk due to severe weather disruptions (Melkonyan et al., 2024; Nimmala, 2024).

Given the growing prominence of climate risks, incorporating diverse climate indicators into credit scoring models has become increasingly important. Table A1 in Appendix A of the supplementary materials summarises existing studies that utilise climate data to predict credit risk. Early studies explored the value of climate data using statistical models. For example, Castro & Garcia (2014) develop a linear regression model to incorporate climate features (temperatures and precipitation) and commodity price volatility to enhance agricultural lending decisions. Similarly, Römer & Musshoff (2018) use a logistic regression model to emphasise the importance of rainfall data for improving default prediction models, particularly in the context of rural finance. In recent years, scholars have increasingly turned to more advanced techniques to model the relationship between climate features and default risk. Using machine learning algorithms, Gao et al. (2023) demonstrate that severe weather events significantly increase default risk among peer-to-peer borrowers. Calabrese et al. (2024) develop an additive Cox proportional hazard model to capture the spatial-temporal characteristics of weather events. Their findings indicate that heavy rainfall and tropical cyclones can significantly increase mortgage default risk. Additionally, Chen et al. (2025) explore the interaction between climate and economic factors using geodetector techniques, which reveals that the combined influence of these factors on credit risk exceeds their individual effects.

*2.2 Integration of text data in credit risk management*

Although traditional default prediction models have primarily relied on structured data, there is a growing body of evidence demonstrating that integrating text data can provide valuable new insights through the leveraging of NLP techniques (Kriebel & Stitz, 2022; Wu et al., 2025). Early studies explored the value of text using conventional text mining approaches. For example, Dorfleitner et al. (2016) derive several textual features from borrowers' loan descriptions, such as spelling errors, text length, and the use of emotional words, and used them to predict funding probability and default risk. As digital technologies flourished, word frequency-based statistical models began to be widely used for processing text data. Pioneering these approaches, Jiang et al. (2018) use latent Dirichlet allocation (LDA) to derive features from borrowers' written descriptions, while Mai et al. (2019) use the term frequency-inverse document frequency (TF-IDF) to extract 20,000 textual representations from the "managerial discussion & analysis" sections of 10-K filings and used them to predict corporate bankruptcy.

Given the breakthroughs in deep learning that have taken place in various fields, researchers have increasingly been exploring the potential of neural NLP models for extracting textual information. Mai et al. (2019) pioneer the use of convolutional neural networks (CNNs), while Matin et al. (2019) utilise recurrent neural networks (RNNs) to predict corporate bankruptcies by analysing audit reports and management statements. Wang et al. (2020) and Kriebel & Stitz (2022) take a different approach by using static word embeddings such as global vectors for word representations (GloVe) as inputs for traditional classifiers like logistic regression and random forests, thereby predicting borrowers' default. More recently, pre-trained language models have revolutionised the field of NLP by offering notable performance enhancements (Vaswani et al., 2017). Among these, transformer models, such as bidirectional encoder representations from transformers (BERT) and the robustly optimised BERT pre-training approach (RoBERTa), are now state-of-the-art across various NLP tasks. While these models are well-known in NLP and AI communities, it was not until 2021 that they were introduced for credit default prediction. Stevenson et al. (2021) first use BERT to derive textual representations from loan officers' assessments, which they then used for bankruptcy prediction. Subsequently, Kriebel & Stitz (2022) demonstrate, using BERT and RoBERTa, that even succinct text descriptions can significantly enhance default prediction. Xia et al. (2023) and Sanz-Guerrero & Arroyo (2024) take an alternative approach, by incorporating predicted default probabilities derived from fine-tuned BERT and RoBERTa as textual features in credit scoring models, showcasing that narrative data contains valuable credit information.

*2.3 Multimodal credit risk modelling*

In practice, the borrower's default risk is influenced by multiple, often complementary, factors that cannot be fully captured by any single information source. With recent advances in data availability and computing technologies, an emerging strand of the literature has begun to adopt multimodal learning frameworks that integrate heterogeneous data sources within a unified modelling architecture. By jointly exploiting information from multiple modalities, these approaches aim to construct richer representations of borrower risk and thereby improve default prediction performance.

Early studies on multimodal credit risk modelling have primarily focused on combining structured credit variables with textual soft information, as discussed in the previous section. This line of work shows that narrative data can provide meaningful signals that enhance discrimination performance, particularly in settings where hard financial information is incomplete, infrequently updated, or slow to reflect changes in credit risk (Wu et al., 2025). In the corporate credit context, most studies use texts from annual and current reports to capture their tones, disclosure styles, and semantic contents that are not fully reflected in financial ratios (Ahmadi et al., 2018; Mai et al., 2019; Matin et al., 2019; Che et al., 2024). More recent studies extend this approach to spoken disclosures, which uses earnings-call transcripts as an additional information channel that complements structured data in assessing credit risk (Yang et al., 2023; Tavakoli et al., 2025). Beyond formal corporate disclosures, researchers increasingly draw on text from the broader information environment. For example, some studies find that lexical and sentiment features extracted from social media and online content can be used to supplement firm fundamentals by providing more timely signals about market perceptions and firm-specific developments (Wang et al., 2020). In addition, user-generated information can help interpret disclosure content and further improve financial distress prediction (Wu et al., 2024). In peer-to-peer lending and mSE settings, text data often take the form of borrower-written loan descriptions or loan-officer assessment narratives. These texts capture soft screening information that is difficult to quantify and have been shown to exhibit strong predictive power, particularly for thin-file borrowers and small firms (Stevenson et al., 2021; Kriebel & Stitz, 2022; Wu et al., 2025).

More recent studies extend multimodal credit risk modelling beyond textual data and demonstrate that default risk can be influenced by multiple information sources that differ in origin, frequency, and coverage. A growing set of studies integrates structured data with different frequencies. For example, some studies combine firm fundamentals and macroeconomic indicators with higher-frequency pricing data to capture both slow-moving solvency conditions and rapidly evolving market assessments of credit risk (Korangi et al., 2023). In parallel, another strand of studies incorporates spatial and sensory information, such as satellite-derived measures and learned spatial embeddings, to proxy for local economic conditions and infrastructure that are relevant to borrower's repayment capacity (Leng et al., 2024; Holvoet et al., 2025). Network-based data have also been increasingly used to capture the propagation of risk across connected entities (Muñoz-Cancino et al., 2023; Che et al., 2024; Zandi et al., 2025). In addition, communication modalities including earnings-call voice data and borrower video data, have been shown to provide complementary signals for predicting credit risk and delinquency (Yang et al., 2023; Chang et al., 2025).

Despite these advances, several important research gaps still remain. First, although a wide range of alternative data sources has been incorporated into credit risk modelling, the role of physical climate risks in influencing loan performance remains underexplored. Existing studies often focus on individual climate events and rely primarily on statistical models to evaluate their effects but not paying much attention to interactions among different climate conditions (Castro & Garcia, 2014; Kaur Brar et al., 2021). Moreover, although prior research shows that climate data and textual data separately add value to credit default prediction, there is limited evidence on the effectiveness of integrating these two modalities within a unified framework. Climate data provide a systematic view of borrowers' exposure to environmental risks, particularly in climate-sensitive sectors such as agriculture and logistics, whereas textual data capture context-specific and qualitative information that structured variables alone cannot convey. Therefore, we anticipate that integrating these complementary sources may significantly improve credit risk assessment. This paper aims to address these gaps and contribute evidence relevant to both academic research and practical applications in business and IS.

3. **Data**

The agricultural loan dataset used in this study was provided by a Chinese bank operating on a national basis. The dataset originally includes 4,500 commercial and industrial loans, primarily granted to agricultural mSEs to allow them to meet immediate operational requirements, such as managing cash flow, financing inventory purchases, or covering unforeseen expenses. These short-term loans are attractive to agricultural mSEs because they can be obtained more quickly than long-term financing options and provide them with the flexibility to manage financial needs without the burden of extended debt commitments. After removing records lacking climate features and textual information, and those with significant missing variables, we have a final sample consisting of 4,172 loans with durations ranging from 1 to 9 months. All the loans are closed – 4,108 loans were fully paid off at maturity or earlier, and 64 borrowers defaulted.[1] Default is defined as 30 days past payment due. Each borrower has 18 nominal and 14 numeric attributes, one text extracted from the loan assessments generated by loan officers, and a final class label (i.e., defaulter or non-defaulter).

*3.1 Structured features*

The original dataset includes 32 standard credit features, which covers borrower's demographics, business details, loan particulars, and so forth. In cases where continuous features are missing, they are filled with the means of the features, while missing values for categorical features are replaced with new categories. Table B1 in Appendix B of the supplementary materials presents the definitions and summary statistics for all standard variables in the dataset. To ensure consistency and comparability of input features across models, we apply the weight of evidence (WoE) method to the features beforehand (Jiang et al., 2019; Wu et al., 2025). The WoE value in each category/bin for a given feature is calculated as the logarithm of the proportion of non-defaulters to the proportion of defaulters in that particular category/bin. A large negative value corresponds to a higher default risk and vice versa. We also employ the information value (IV) and variance inflation factor (VIF) measures to remove redundancy, reduce the multicollinearity among WoE-encoded features, and select an appropriate subset of features for inclusion in the models (Wu et al., 2025). We calculate the IV as the weighted sum of the WoE values for each category/bin of the feature, which serves to represent the feature's predictive power. We retain WoE-encoded features with an IV greater than 0.01 and less than 0.50 (Anderson, 2007). Also, to detect multicollinearity issues among WoE-encoded features, we only retain those with a VIF of 10 or less. These preprocessing steps allow us to identify 21 structured features suitable for our subsequent modelling.

---

[1] We note that the dataset is characterised by a severely imbalanced ratio of 64:1 (non-defaulters: defaulters). However, this is quite common among Chinese lenders. For example, according to the China Banking and Insurance Regulatory Commission, the average non-performing loan ratio for the Chinese banking sector was 1.56% in 2024.

*3.2 Climate features*

We first collect daily meteorological data from 440 prefecture-level meteorological stations across China, each monitored by the National Oceanic and Atmospheric Administration (Liu et al., 2025).[2] Then we compute monthly climate risk indices from the corresponding daily measures and adjust for seasonal variations and regional differences. Using geographic coordinates, we identify the meteorological station nearest to each mSE's headquarters[3] location and derive four climate risk factors, drought, water-logging by rain, high-temperature, and cryogenic freezing risks, for each loan. Each risk factor is measured for the 12 months preceding the loan's start date, which yields 12 monthly observations per loan. Consequently, for each type of climate risk, our sample of 4,172 loans produces 50,064 loan-month observations. This methodology follows prior research such as Pelka et al. (2015) and Römer & Musshoff (2018). The following subsections provide a detailed description of the methods used to derive these climate features.

*3.2.1 Drought risk*

We first calculate the daily drought ($D_d$) index for each meteorological station to quantify the daily intensity of drought conditions, based on the corresponding level of the standardised precipitation index (*SPI*) (Wang et al., 2018). Due to space constraints, detailed information on the computation of the *SPI* is provided in Appendix C of the supplementary materials.

---

[2] China is located in the East Asian monsoon region, which has a distinctive monsoon climate. In spring and summer, floods and droughts occur frequently in China, while autumn and winter are characterised by cold air outbreaks and cold waves. Certain regions experience an increased number of snowfall days, deeper snow cover, and more significant impacts, while the number and landfall frequency of typhoons are relatively low. Based on an analysis of climate risk exposure characteristics specific to China's context, this study ultimately identifies drought, water-logging by rain, high-temperature, and cryogenic freezing as the primary climatic events (Zhou et al., 2009; Li et al., 2022; Li et al., 2023; Zhang et al., 2023).
[3] In this study, we do not distinguish between different types of mSE geographical locations (e.g., headquarters versus business locations). In our dataset, the borrowers are agricultural mSEs that typically conduct their core production or business activities at a limited number of sites, often located in close proximity to their registered locations. This makes the registered address an appropriate basis for linking each borrower to the nearest meteorological station when constructing climate risk measures (Kiggundu, 2002; Choongo et al., 2021).

$$D_d = \begin{cases} 0, & SPI \geq -1 \\ SPI + 1, & -1.5 \leq SPI < -1 \\ 2 \times SPI + 2.5, & -2 \leq SPI < -1.5 \\ 3 \times SPI + 4.5, & SPI < -2 \end{cases} \quad (1)$$

We then use Equation (2) to calculate the monthly drought index ($DI$) as the measure of drought risk for each station. In this formulation, $Day$ denotes the number of days in the month with an average temperature above 0 °C, and $D_{d,i}$ represents the daily drought index on day $i$. Here, $a_i$ is the regional weight assigned to the station, which takes a value of 0 for stations located on the Tibetan Plateau and in the central and western parts of Northwest China, 0.6 for stations in the eastern part of Northwest China and the southwestern region, and 1 for all other regions. The term $b_i$ denotes the monthly weight, set to 1.5 for May through September, 1 for March, April, October, and November, and 0.5 for December, January, and February.

$$DI = \sum_{i=1}^{Day} D_{d,i} \times a_i \times b_i \quad (2)$$

*3.2.2 Water-logging by rain risk*

We compute the monthly water-logging by rain ($WLR$) index, which serves as the measure of water-logging risk, for each meteorological station based on the daily precipitation data at the prefecture level (Bonsal & Regier, 2007; Wang et al., 2018). We first calculate the daily precipitation index ($R_d$) using Equation (3), where $P$ denotes the daily precipitation and $n$ represents the $n$-th consecutive day of rainfall.

$$R_d = \begin{cases} 0, & P < 50mm \\ n^{\frac{1}{2}}, & 50mm \leq P < 100mm \\ 2n^{\frac{1}{2}}, & 100mm \leq P < 200mm \\ 3n^{\frac{1}{2}}, & 200mm \leq P \end{cases} \quad (3)$$

We then calculate the monthly $WLR$ for each station according to Equation (4), where $Day$ denotes the total number of days in the month and $c_i$ is the monthly weighting coefficient (2 for June, July, and August, and 1 for all other months).

$$WLR = \frac{\sum_{i=1}^{Day} R_{d,i}}{Day} \times c_i \quad (4)$$

*3.2.3 High-temperature risk*

We calculate the monthly high-temperature ($HT$) index as the measure of high-temperature risk for each meteorological station using daily maximum and minimum temperature, and duration of high temperatures at the prefecture level (Alexander et al., 2006; Wang et al., 2018). First, we calculate the daily maximum temperature index ($T_g$) and the daily minimum temperature index ($T_d$) as follows:

$$T_g = \begin{cases} 1, & 35°C \leq T_{max} < 37°C \\ 2, & 37°C \leq T_{max} < 40°C \\ 3, & 40°C \leq T_{max} \end{cases} \quad (5)$$

$$T_d = \begin{cases} 1, & 25°C \leq T_{min} < 28°C \\ 2, & 28°C \leq T_{min} < 30°C \\ 3, & 30°C \leq T_{min} \end{cases} \quad (6)$$

We then calculate the monthly $HT$ using Equation (7), where $Day$ denotes the number of days per month, $D_{g,i}$ is the number of consecutive days on which the daily maximum temperature is at least 35°C, and $D_{d,i}$ is the number of consecutive days on which the daily minimum temperature is at least 25°C.

$$HT = \frac{\sum_{i=1}^{Day} T_{g,i} \times (D_{g,i})^{\frac{1}{2}} + \sum_{i=1}^{Day} T_{d,i} \times (D_{d,i})^{\frac{1}{2}}}{Day} \quad (7)$$

*3.2.4 Cryogenic freezing risk*

To calculate the monthly cryogenic freezing ($CF$) index as the measure of cryogenic freezing risk, we incorporate temperature variability and the number of snow days over a five-day timescale (Wang et al., 2018). First, we define index $a$ based on the standard deviation ($\sigma$) of the average temperature anomaly over each five-day period as follows. Let $t$ represent the average temperature for the five-day window, and $\bar{t}$ denote the corresponding 20-year (2001-2020) climatological average temperature for the same period.

$$a = \begin{cases} 0, & (t - \bar{t}) > -1\sigma \\ 1, & -2\sigma < (t - \bar{t}) \leq -1\sigma \\ 2, & -3\sigma < (t - \bar{t}) \leq -2\sigma \\ 3, & (t - \bar{t}) \leq -3\sigma \end{cases} \quad (8)$$

Then, we calculate the daily cryogenic freezing index, $I_c$, for each five-day period using Equation (9). Here, $c_i$ is the weighting coefficient reflecting snowfall intensity, defined as $1 + Day/10$, where $Day$ is the number of snow days in the five-day window. The term $d_i$ denotes the regional weight assigned to each station, which takes a value of 0.5 for southern China and 1 for northern China.

$$I_c = a \times \left|\frac{t - \bar{t}}{\sigma}\right| \times c_i \times d_i \quad (9)$$

Finally, using Equation (10), we can obtain the monthly $CF$ by summing the six five-day indices within each month:

$$CF = \sum_{i=1}^{6} I_c(i) \quad (10)$$

Because the effects of cryogenic freezing are most severe during winter, we apply seasonal weights: December receives a weight of 1, January and February receive a weight of 2, and all other months receive a weight of 0.5, which reflects their comparatively lower exposure to freeze-related impacts.

After obtaining the four types of climate risk features, we report their definitions and summary statistics in Table 1. Figure 1 further displays the spatial distribution of the yearly climate risk indices at the meteorological-station level, overlaid with the locations of individual loans, where red and blue dots denote defaulters and non-defaulters, respectively.

**Table 1** Definitions and statistical information for climate features

| Climate feature | Definition | Summary statistics | | | | |
|---|---|---|---|---|---|---|
| | | *Obs* | *Mean* | *SD* | *Min* | *Max* |
| Drought risk | Measures monthly dryness severity based on precipitation anomalies and temperature-adjusted drought indices. | 50,064 | 9.74 | 1.30 | 0.00 | 10.00 |
| Water-logging by rain risk | Measures monthly excess-rainfall severity using accumulated and consecutive-day precipitation. | 50,064 | 0.88 | 1.39 | 0.00 | 10.00 |
| High-temperature risk | Measures monthly heat stress using daily maximum/minimum temperatures and high-temperature duration. | 50,064 | 0.02 | 0.22 | 0.00 | 8.19 |
| Cryogenic freezing risk | Measures monthly cold-stress severity based on temperature anomalies and snow-related freezing conditions. | 50,064 | 0.37 | 1.17 | 0.00 | 9.19 |

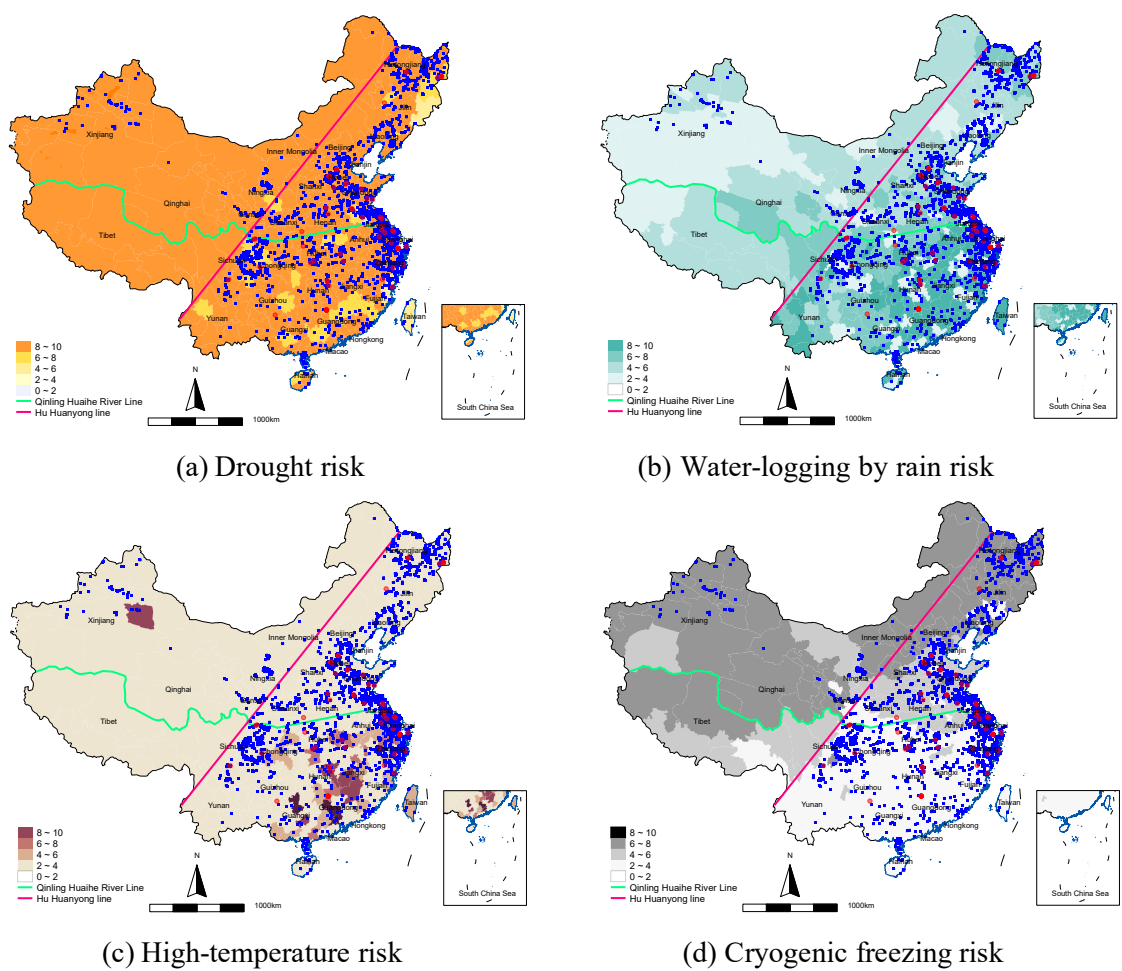

**Figure 1** Spatial distribution of the yearly climate risks across China, where the red (blue) dots denote the locations of defaulters (non-defaulters)

*3.3 Text description*

In many cases, mSEs lack a credit history and struggle to provide necessary and reliable financial information (Katchova & Barry, 2005). To gain a comprehensive understanding of an mSE's operational and financial situation, loan officers from the lender usually conduct site visits (Stevenson et al., 2021). These visits allow the lenders to verify the information provided in the loan applications and gather first-hand details about the borrower's business activities, day-to-day operations, and repayment commitment, among other things (Wu et al., 2025). During these visits, loan officers record statements from borrowers and then provide their own assessments of various aspects, thereby assisting the loan underwriters to evaluate the potential credit risk of the borrowers and make informed decisions on loan approvals. The loan underwriters would thoroughly review the documents provided, alongside the borrower's applications, financial information, and any available credit histories, to make these decisions. In our dataset, the loan officer's assessment typically evaluates the borrower from three perspectives: an overall impression of the borrower, a review of the borrower's credit history, and the borrower's repayment intentions. To ensure the text is clear for our analysis, we remove redundant white spaces in the text and add a full stop at the end of each complete sentence, steps which can enhance a text's readability. After pre-processing, the minimum, mean, and maximum word counts of the texts are 15, 107, and 326, respectively, with a standard deviation of 38. Below, we present some typical extracts from the textual loan assessments, which have been translated into English from the original Chinese. These extracts have been fully anonymised, with all quantitative values masked.

Extract 1: *"The borrower has actively cooperated with the investigation, and the loan officer holds a favourable impression of him. The borrower currently maintains a good credit rating, and this is his fourth loan, with timely repayments having been made for the previous three. The borrower has a strong social network, fosters harmonious relationships with neighbours, and is regarded as honest".*

Extract 2: *"The borrower has shown reasonable cooperation with the pre-loan investigation and demonstrates diligence and perseverance, which indicates that he is a trustworthy individual. The borrower maintains harmonious relationships with his neighbours. Based on an investigation of the borrower's family situation and farm operations, the loan officer found that the borrower was well-versed in livestock management and had a stable cash flow. Additionally, several sows on the farm are currently close to farrowing".*

Extract 3: *"The borrower was highly cooperative with the loan officer during the pre-loan investigation, and exhibited a composed and gracious demeanour. The loan officer indicated that the borrower was hardworking and resilient, maintaining harmonious and amicable relationships with both his spouse and neighbours. The loan officer assigned the borrower a credit rating score of ??? points, classifying him as an AA-level client. The borrower stated that he had undertaken a landscaping project in Suzhou and required a loan for some advance funding".*

4. **Models**

The main objective of our study is to develop a multimodal framework that integrates multiple data sources for credit risk prediction. Specifically, our prediction models are constructed using three types of features: structured credit data, climate panel data, and unstructured textual narratives. This design yields seven distinct model specifications, corresponding to different combinations of the input modalities, as summarised in Table 2. To ensure comparability across specifications, we adopt a simple multilayer perceptron as the benchmark classifier for default prediction (Stevenson et al., 2021; Wu et al., 2025). Each unimodal model is trained exclusively on a single data modality. For example, when the model relies solely on text data, the structured and climate inputs are deactivated, and the prediction is based only on textual representations. In contrast, multimodal models jointly exploit multiple data channels, which enables the learning of interactions across modalities. During training, all modality-specific components are optimised simultaneously. For example, when all three data sources are incorporated, structured features are concatenated with the latent representations extracted from the text and climate models. The resulting combined representation is then passed through a dense layer to produce the final binary classification. Figure 2 illustrates the architecture of the full multimodal model based on structured, climate, and text data. Sections 4.1-4.3 then briefly describe the LSTM, GRU, and transformer architectures employed in our empirical analysis. While our empirical analysis focuses on LSTM-, GRU-, and transformer-based architectures, the proposed framework is flexible and can be readily extended to other data types and models.

**Table 2** Input features for various default prediction models

| Modality/Data type | Structured data | Climate data | Text data |
|---|---|---|---|
| Structured | 7 continuous, 14 categorical | - | - |
| Climate | - | 4 factors × 12 months | - |
| Text | - | - | token embeddings |
| Structured+Climate | 7 continuous, 14 categorical | 4 factors × 12 months | - |
| Structured+Text | 7 continuous, 14 categorical | - | token embeddings |
| Climate+Text | - | 4 factors × 12 months | token embeddings |
| Structured+Climate+Text | 7 continuous, 14 categorical | 4 factors × 12 months | token embeddings |

**Notes:** Token embeddings are 300-dimensional for fastText-based models and 768-dimensional for BERT-based models.

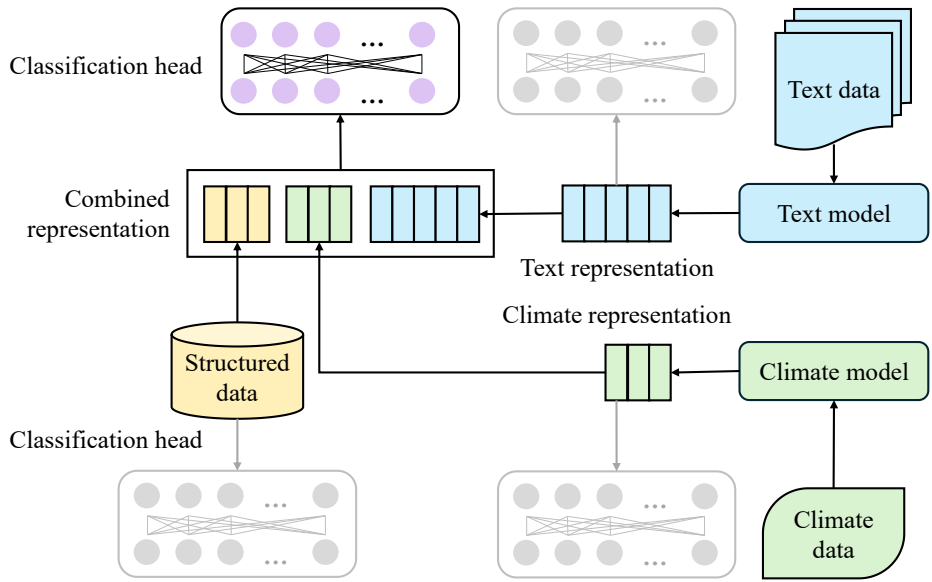

**Figure 2** Multimodal model architecture (structured + climate + text)

*4.1 Long short-term memory (LSTM)*

RNNs were initially developed to process sequential data by iterating through each element in a sequence and using a hidden state to store information. As a result, the output of an RNN depends on both the current input and the previous hidden state through a recurrence mechanism. However, as the number of time steps increases, conventional RNNs often face challenges during training due to the gradients either becoming too small (vanishing gradients) or too large (exploding gradients), which hinders the model's ability to learn long dependencies effectively.

LSTM features better memory capability to address these limitations by deploying memory cells with several gates in their hidden layers (Hochreiter & Schmidhuber, 1997). Each LSTM cell has three gates: a forget gate, an input gate, and an output gate. The forget gate uses a sigmoid function to decide which parts of the long-term cell state should be passed through based on the priority and usefulness of the information for prediction. The input gate determines how much of the new information from the current input, and the previous short-term hidden state should be added to the cell state. It uses a combination of a sigmoid function to determine which values to update and a tanh function to generate candidate values for addition. The output gate, governed by a sigmoid activation, decides which parts of the cell state contribute to the current output, filtered through a tanh activation to ensure a smooth range of output values. These gating mechanisms enable the LSTM to retain or discard information across multiple time steps, effectively addressing the vanishing and exploding gradient problems. This enhanced memory capability has made LSTM a cornerstone in applications such as machine translation, speech recognition, and time-series forecasting (Tavakoli et al., 2023). In this study, we configure the LSTM models with 128 hidden units and determine the optimal number of layers, learning rate, and batch size based on the best performance on the validation sample. To process the text data, we use fastText to derive 300-dimensional word embedding for each token in a sentence; these word embeddings are then fed into the LSTM networks. Finally, we add a dense layer on top of the LSTM backbone to perform the downstream binary classification.

*4.2 Gated recurrent unit (GRU)*

As another variant of RNNs, the GRU provides a more straightforward architecture than LSTM to address challenges of vanishing/exploding gradients through the use of two gates: a reset gate and an update gate (Cho et al., 2014). The reset gate uses a sigmoid function to determine the extent to which previous memory should be combined with the current input for loading into the new memory. This mechanism enables the GRU to selectively incorporate past information based on its relevance to the current context. The update gate, also controlled by a sigmoid activation, determines the proportion of existing memory to retain and how much of the new memory to incorporate. This unified gating structure effectively balances the trade-off between preserving past information and updating with current inputs, enabling the GRU to learn long-term dependencies efficiently (Lynn et al., 2019). Due to its simpler architecture and fewer parameters compared to LSTM, GRU trains faster and requires less computational power. In this study, we configure the GRU models with 128 hidden units and select the optimal number of layers, learning rate, and batch size[4] based on their performance on the validation sample. As with LSTM, to process the text data, we utilise fastText to generate 300-dimensional word embedding for each token in the sentence, and these word embeddings are then fed into the GRU networks. On top of the GRU backbone, we add a dense layer for the downstream binary classification.

---

[4] For the LSTM and GRU models, the hyperparameter search explores the number of layers [2, 3], learning rate [2e-5, 1e-5, 1e-4, 1e-3], and batch size [16, 32]. For the transformer models, the search focuses only on the learning rate [2e-5, 1e-5, 1e-4, 1e-3] and batch size [16, 32].

*4.3 Transformers*

Unlike LSTM and GRUs, transformers utilise a self-attention mechanism to model relationships among all elements in a sequence simultaneously, rather than processing inputs sequentially. Self-attention assigns weights to each element in the input sequence based on its relevance to all other elements, which enables the model to capture both short- and long-term dependencies. To enhance this capability, transformers employ multiple attention heads, each of which focuses on different aspect of the input and jointly learns diverse and complex relationships. Since self-attention alone does not encode the order of sequence elements, transformers incorporate positional encodings into the input embeddings to provide explicit information about token positions. This design removes the need for recurrent computation, which allows all elements in the sequence to be processed in parallel and significantly improves computational efficiency. A standard transformer architecture consists of an encoder and a decoder. The encoder processes the input sequence and produces a higher-level representation through stacked self-attention and feed-forward layers. The decoder then generates the output sequence by applying self-attention to previously generated outputs and cross-attention to the encoder's representations. The resulting representations are passed through a final feed-forward layer to produce the model output (Vaswani et al., 2017). While the full encoder-decoder structure is commonly used for sequence-to-sequence tasks, many classification applications rely only on the encoder to extract high-level representations, which are then passed to a downstream classifier.

Transformers were originally developed to model sequential data, which makes them well suited for classification tasks with panel or time-series inputs. In this study, we adapt the standard transformer by using only the encoder part to construct representations of the climate features, following Korangi et al. (2023), as shown in Figure 3(a). The encoder captures temporal relationships across time periods and transforms the input sequence into a set of latent representations that are appropriate for classification. Figure 3(b) shows an example with twelve time periods of panel data, where the climate features at each time step are mapped to a fixed-dimensional representation. The dimensionality of these representations (the model size) is a predefined hyperparameter of the transformer. As our problem is a binary classification task, we subsequently pass the encoder output to a dense layer to produce the final predictions, and optimise the entire model by selecting the optimal learning rate and batch size based on validation performance (Wu et al., 2025). In our implementation, the transformer encoder uses a model size of 128, eight attention heads, and three encoder layers, with a feed-forward layer dimension of 256.

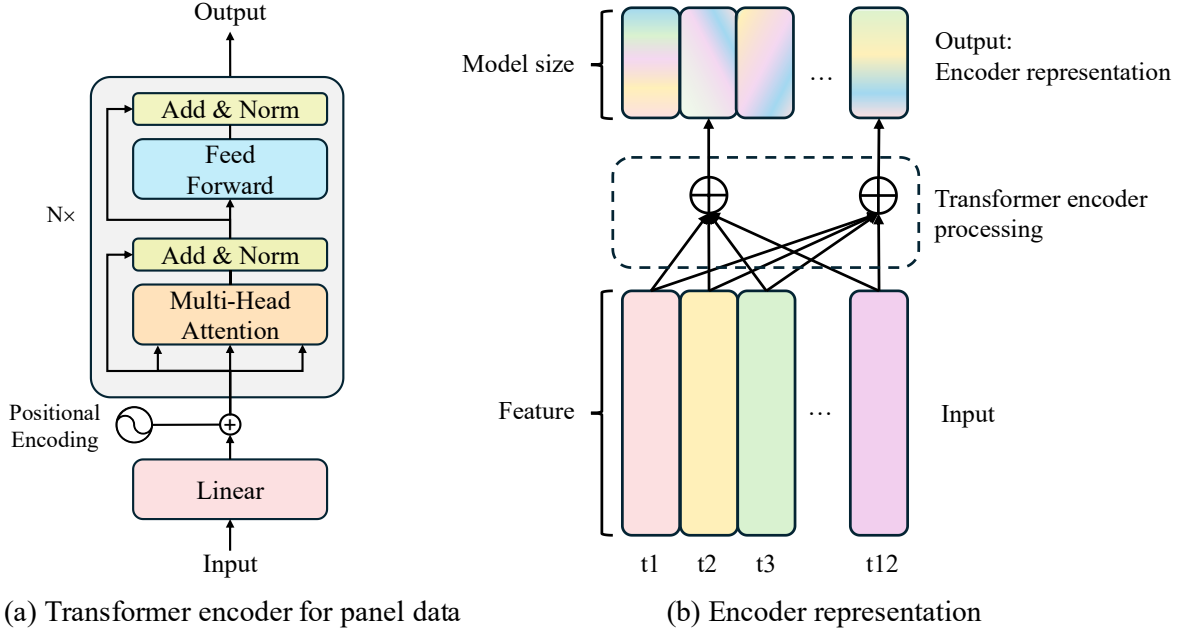

(a) Transformer encoder for panel data      (b) Encoder representation

**Figure 3** Transformer encoder architecture and representation

In the NLP context, text naturally exhibits a sequence-like structure governed by grammatical rules and contextual dependencies. Each token in a sentence can be viewed as an ordered element analogous to a time step and can be mapped into an embedding that captures attributes such as spelling, meaning, and other linguistic features. To process the text data in our study, we deploy BERT to extract contextualised representations (Stevenson et al., 2021; Kriebel & Stitz, 2022). BERT is an encoder-only transformer model developed by Google AI and pre-trained using masked language modelling and next-sentence prediction objectives (Devlin et al., 2019). By learning bidirectional representations from both left and right contexts, BERT overcomes the limitations of unidirectional language models and provides rich contextual embeddings. Owing to its pre-training on large-scale corpora and large model capacity with hundreds of millions of parameters, BERT has demonstrated strong performance across a wide range of NLP tasks. Given that our text data are in Chinese, we use multilingual BERT trained on word pieces over 100+ lexica to derive the text representation. On top of the BERT backbone, we add a dense layer for the downstream binary classification task and the entire model is fine-tuned by selecting the optimal learning rate and batch size based on validation performance.

Given the computational complexity of transformer-based models, we adopt a hybrid strategy for multimodal learning. Specifically, we first train unimodal transformer models independently on climate and text data, respectively, and save the best-performing specifications. During multimodal training, the parameters of these pre-trained transformer components are frozen, and only the dense fusion layer is updated. This approach focuses learning on the integration of the combined representations while maintaining stable and efficient optimisation (Stevenson et al., 2021; Korangi et al., 2023).

## 5. Experimental Design

This section describes the loss function used to optimise the models during training, outlines the evaluation metrics used for assessing the prediction performance, and introduces the methods used for the interpretation of the models.

*5.1 Loss function*

We divide our dataset into three distinct subsets – training, validation, and test – following the approach taken by Kriebel & Stitz (2022). The training set is used to train the models, the validation set is used to choose optimal model hyperparameters, and the test set is used to estimate the out-of-sample prediction results. To construct these sets, we use the stratified random sampling method to select 70% of the samples as the training set, 20% of the training set as the validation set, and the remaining 30% as the test set.

Given the nature of our binary-label classification task, we chose to base our models on the binary cross entropy (BCE) loss. The BCE loss for the entire set of observations is denoted by $L(y, \hat{y})$, whose calculation is given in Equation (11) below, where $N$ represents the batch size, $y_i$ is the actual label (0 for non-defaulters and 1 for defaulters), and $\hat{y}_i$ is the predicted probability of default for observation $i$. We run training for multiple epochs and select hyperparameters that minimise the BCE loss on the validation set. Models with these optimised hyperparameters are saved and utilised to predict the probabilities of default for the test set.

$$L(y, \hat{y}) = -\frac{1}{N} \sum_{i=1}^{N} (y_i \cdot \log(\hat{y}_i) + (1 - y_i) \cdot \log(1 - \hat{y}_i)) \qquad (11)$$

*5.2 Performance evaluation*

Discrimination performance refers to the ability to distinguish between negative and positive classes. While several measures exist for gauging discrimination performance, prior research suggests that using multiple measures provides a more comprehensive assessment (Lessmann et al., 2015). In this study, we prefer aggregate performance metrics to point metrics for several reasons. Aggregate performance metrics offer a threshold-independent evaluation of a classification model's performance, offering a holistic assessment of the model's ability to discriminate between classes over the entire range of thresholds. However, point metrics are calculated at a specific decision threshold, which may not reflect a model's overall performance and can be chosen arbitrarily (Junuthula et al., 2016; Bao et al., 2022). This means that the use of point metrics for model comparison can be problematic due to the significant variation in performance at different thresholds.

Therefore, we evaluate the performance of our credit scoring models using three widely used metrics: the area under the receiver operating characteristic (ROC) curve (AUC), the Kolmogorov-Smirnov (KS) statistic, and the H-measure. The ROC curve is a graphical representation of a classification model's predictive accuracy over a range of threshold values. The curve plots the true positive rate (i.e., sensitivity) against the false positive rate (i.e. 1 – specificity) for each threshold value of a given model (Wu & Li, 2021). The KS statistic is defined as the maximum vertical distance between the empirical cumulative distribution functions of the false positive rate and the true positive rate, and can be used to assess the correctness of categorical predictions. Proposed by Hand (2009), the H-measure avoids the deficiency of the AUC, in that it uses different misclassification cost distributions for different classifiers, by specifying a preset severity ratio for assessing the impact of the misclassification cost between negative and positive instances. In this study, we adopt the standard severity ratio, which is the inverse of the relative class frequency (Chen et al., 2024). Typically, the larger the AUC, KS, and H-measure, the better the performance of a prediction model.

To estimate the discrimination performance of each model, we initially employ five random seeds to conduct multiple runs of all experiments. We then apply the bootstrap resampling method, conducting 1,000 resamples for each of the five random seeds (Berg-Kirkpatrick et al., 2012; Deldjoo, 2023; Katsafados et al., 2024). This yields a total of 5,000 performance estimates for the test set, thereby reducing randomness and enhancing the reliability and stability of the results. The discrimination performance results (mean and its 95% confidence interval) reported later are all based on the 5,000 estimates.

*5.3 Model interpretability*

In the business context, the interpretation of a model's results plays an essential role in data-driven decision-making. However, deep learning models are often considered black-boxes and are not always well understood, due to their complexity. In our study, to provide a better understanding of how different data modalities influence the prediction results of credit scoring models, we use SHAP to explain the relative importance and the model's temporal dependence (Korangi et al., 2023). SHAP is an interpretability framework based on the Shapley value concept from game theory, which assigns fair payouts to players depending on their contributions to the total gain, considering all possible coalitions (Shapley, 1953). In the context of credit scoring, SHAP treats feature values as players and coalitions as subsets of features, enabling it to measure the marginal contribution of each feature to the model's prediction (Lundberg & Lee, 2017). By averaging these marginal contributions across all possible feature subsets and orderings, the SHAP values provide a robust quantification of the importance of individual features.

Compared to other explainable AI models, SHAP has three critical properties: local accuracy, missingness, and consistency, which ensure the fairness and reliability of its interpretability. Local accuracy requires that the sum of all feature contributions equals the model's output for any given prediction. This property guarantees that SHAP provides a complete decomposition of the model's behaviour. Missingness assigns a contribution value of zero to any feature that does not influence the model's output, which reinforces the method's robustness by appropriately excluding irrelevant features from the explanation. Consistency states that, if the contribution of a feature increases in the model (e.g., due to a change in the model's parameters or input data), its SHAP value will not decrease. These theoretical guarantees make SHAP a unique and widely applicable tool for interpreting credit scoring models (Chen et al., 2024).

## 6. Empirical Results

In this section, we present three sets of results. First, we evaluate the performance of unimodal models trained on individual data modalities to assess their effectiveness in default prediction independently. Next, we showcase the outcomes of multimodal learning, which incorporates various combinations of the data modalities to examine their influence on predictive performance. Finally, we conduct an interpretability analysis of our best-performing model architecture, focusing on how different types of data, particularly climate risk factors, influence prediction outcomes.

*6.1 Unimodal models*

First, we consider each individual data modality as an input to build the unimodal models. Table 3 presents the discrimination performance of different deep learning models.[5] [6] In the first three rows of each part of the table, we show the results of using structured-only (S), climate-only (C), and text-only (T) data, respectively.

**Table 3** Model performance across different data modalities

| Model | Modality | AUC | KS | H-measure |
|---|---|---|---|---|
| LSTM | Structured | 0.609 (0.607, 0.610) | 0.301 (0.299, 0.302) | 0.157 (0.155, 0.159) |
| | Climate | 0.628 (0.626, 0.629) | 0.304 (0.301, 0.306) | 0.174 (0.172, 0.176) |
| | Text | 0.613 (0.611, 0.615) | 0.293 (0.290, 0.296) | 0.148 (0.146, 0.150) |
| | Structured+Climate | 0.708 (0.706, 0.710) | **0.450 (0.447, 0.452)** | **0.269 (0.266, 0.271)** |
| | Structured+Text | 0.667 (0.665, 0.669) | 0.385 (0.382, 0.387) | 0.175 (0.173, 0.177) |
| | Climate+Text | 0.683 (0.681, 0.684) | 0.394 (0.392, 0.397) | 0.210 (0.208, 0.212) |
| | Structured+Climate+Text | **0.716 (0.715, 0.718)** | 0.423 (0.420, 0.425) | 0.247 (0.245, 0.249) |
| GRU | Structured | 0.609 (0.607, 0.610) | 0.301 (0.299, 0.302) | 0.157 (0.155, 0.159) |
| | Climate | 0.658 (0.656, 0.660) | 0.332 (0.329, 0.334) | 0.189 (0.187, 0.192) |
| | Text | 0.660 (0.658, 0.662) | 0.345 (0.342, 0.347) | 0.211 (0.209, 0.213) |
| | Structured+Climate | 0.708 (0.707, 0.709) | 0.406 (0.404, 0.408) | **0.235 (0.233, 0.237)** |
| | Structured+Text | 0.676 (0.674, 0.678) | 0.375 (0.372, 0.378) | 0.200 (0.198, 0.202) |
| | Climate+Text | 0.675 (0.673, 0.677) | 0.371 (0.368, 0.375) | 0.221 (0.218, 0.224) |
| | Structured+Climate+Text | **0.709 (0.708, 0.710)** | **0.452 (0.450, 0.455)** | 0.221 (0.219, 0.223) |
| Transformer | Structured | 0.609 (0.607, 0.610) | 0.301 (0.299, 0.302) | 0.157 (0.155, 0.159) |
| | Climate | 0.636 (0.633, 0.638) | 0.308 (0.305, 0.312) | 0.182 (0.179, 0.185) |
| | Text | 0.661 (0.658, 0.663) | 0.371 (0.368, 0.375) | 0.244 (0.241, 0.246) |
| | Structured+Climate | 0.697 (0.695, 0.699) | 0.396 (0.393, 0.399) | 0.230 (0.227, 0.232) |
| | Structured+Text | 0.681 (0.679, 0.684) | 0.381 (0.378, 0.384) | 0.233 (0.231, 0.236) |
| | Climate+Text | 0.717 (0.715, 0.719) | 0.433 (0.430, 0.436) | 0.294 (0.291, 0.296) |
| | Structured+Climate+Text | **0.747 (0.745, 0.748)** | **0.464 (0.461, 0.467)** | **0.306 (0.304, 0.309)** |

**Notes**: This table presents the discrimination performance, calculated using the AUC, KS, and H-measure (mean and its 95% confidence interval), of three model architectures: LSTM, GRU, and transformer, used for predicting the default risk of borrowers.

The average AUC, KS, and H-measure for the model relying solely on structured data (S) are 0.609, 0.301, and 0.157, respectively. As for the models trained on the climate-only data (C), all models outperform the structured-only models and achieve impressive results. Given that our loan data is based on the agricultural setting, we can likely attribute this superior performance to the high relevance of climate risk factors such as droughts, floods, and temperature extremes to borrowers' repayment capabilities. For example, prolonged droughts or heavy rainfall can directly impact crop yields and livestock productivity, which reduces borrowers' incomes and ability to meet loan repayment obligations. Interestingly, while transformer models generally exhibit superior performance in unimodal analyses, they are slightly outperformed by the GRU model, which achieves average AUC, KS, and H-measure values of 0.658, 0.332, and 0.189, respectively. These results may stem from the relatively lower complexity of the climate data, which prevents the full utilisation of the transformers' advanced self-attention mechanisms.

---

[5] We also include prediction results using alternative classifiers including logistic regression and extreme gradient boosting as robustness checks to supplement our main experiments. The related results, which support our main findings, are reported in Table D1 in Appendix D.

[6] As a robustness check for the sample splitting, we additionally implemented the 5-fold cross validation. The results reported in Table D2 in Appendix D are consistent with our initial findings and further confirm the stability and effectiveness of our data-splitting approach.

When we examine the performance of the text-only models (T), we find that most of the prediction results outperform those generated by the structured-only models (S). Among the three models, the transformer model achieves the best performance, with average AUC, KS, and H-measure values of 0.661, 0.371, and 0.224, respectively. This superior performance can be attributed to the transformer's ability to process and understand the contextual and semantic meaning within text data, which enables it to extract meaningful patterns that enhance the predictive accuracy. Compared to the results of the structured-only models, these results suggest that loan officers' written assessments may contain valuable insights that reflect borrowers' repayment intentions and operational capabilities. Integrating this information into prediction models may help to capture subtle borrower characteristics that structured data may ignore.

Overall, we find that the models utilising the climate and text data respectively outperform those built on the structured data, which indicates the significant predictive value of alternative data for assessing borrower's default risk. These findings further highlight the potential for integrating multimodal data to build more robust and accurate prediction models, thereby supporting lending decision-making.

*6.2 Multimodal models*

The multimodal model is designed to utilise data from different modalities using the architecture proposed in Figure 2. This framework enables the model to capture diverse patterns across modalities while maintaining flexibility with respect to varying data dimensions and structures. To systematically assess the influence of data fusion on default prediction performance, we first explore three pairwise combinations of the modalities: structured + climate (S+C), structured + text (S+T), and climate + text (C+T). We then integrate all three modalities to examine the performance of the full multimodal integration (S+C+T). Table 3 presents the discrimination performance of the different multimodal models.

Table 3 shows that the combination of structured data and climate data (S+C) yields robust performance across all model architectures. In this setting, the RNN-based models achieve particularly strong results, significant outperforming their corresponding unimodal benchmarks. For models trained on structured and text data (S+T), we observe a similar pattern where multimodal specifications outperform both structured-only (S) and text-only (T) models. Among the three architectures, the transformer consistently achieves the strongest performance. This advantage can be attributed to its superior ability to model semantic and contextual information in textual data. These findings suggest that contextualised language models (e.g., BERT) are more effective at extracting informative signals from text than static word embedding approaches (e.g., fastText), in line with previous findings (Wu et al., 2025). When considering the integration of climate and text data (C+T), model performance again improves relative to the unimodal configurations. Consistent with earlier results, the transformer achieves the highest average AUC, KS, and H-measure values (0.717, 0.433, and 0.294, respectively). In contrast, while the RNN-based models also benefit from this multimodal integration, their performance remains comparatively lower.

We further examine models that integrate all three data modalities (S+C+T). These fully multimodal models achieve the strongest overall performance across architectures in most cases. Notably, the transformer-based model achieves the highest average AUC, KS, and H-measure values (0.740, 0.464, and 0.306, respectively) among all reported results. This superior performance demonstrates the transformer's ability to effectively integrate heterogeneous information from structured credit data, climate panel data, and unstructured text. The consistent performance gains observed when incorporating three modalities, relative to models using one or two modalities, further highlights the importance of information fusion for capturing diverse drivers of default risk.

Overall, these findings suggest that integrating multiple data modalities significantly enhances predictive accuracy and enables a more comprehensive assessment of borrower default risk. This result is particularly salient in settings where financial outcomes are jointly influenced by borrower-specific characteristics and external systemic factors, such as climate-related risks.

### 6.3 Interpretability of the architecture

Although our multimodal models demonstrate strong predictive power, one key challenge lies in the interpretation of the factors contributing to these predictions. To evaluate the influence of each data modality on the multimodal model and the shifts in relative modal importance, we first review the correlation between the different sets of model outputs (Stevenson et al., 2021). Given that the transformer model achieves the best performance in Table 3, the following analysis focuses on this model specification. In Table 4, we opt for the Spearman rank definition of correlation as it is non-parametric and allows us to gain insights into whether the ordering of the probabilities (and, hence, the risk ranking of applicants) has fundamentally changed in the aggregated test sets.

**Table 4** Spearman's rank correlations between predicted default probabilities (test set)

| Modality | S | C | T | S+C | S+T | C+T | S+C+T |
|---|---|---|---|---|---|---|---|
| S | 1.000 | | | | | | |
| C | -0.007 | 1.000 | | | | | |
| T | 0.017 | 0.055 | 1.000 | | | | |

| | | | | | | |
|---|---|---|---|---|---|---|
| S+C | 0.477 | 0.808 | 0.068 | 1.000 | | |
| S+T | 0.842 | -0.063 | 0.055 | 0.373 | 1.000 | |
| C+T | 0.008 | 0.891 | -0.021 | 0.750 | 0.041 | 1.000 |
| S+C+T | 0.483 | 0.807 | 0.025 | 0.936 | 0.400 | 0.816 | 1.000 |

The results in Table 4 suggest little agreement among the structured-only, climate-only, and text-only models as they are weakly correlated, with coefficients of -0.007, 0.017, and 0.055, respectively. This weak alignment shows that each modality captures distinct aspects of the borrowers' credit characteristics, which emphasises their complementary roles in default prediction. When we examine the correlations between the results of multimodal models and those of the unimodal models, we can observe that climate data serves as a key driver in the information fusion. For example, the S+C and S+C+T models are highly correlated with the climate-only (C) model, which shows correlation coefficients of 0.808 and 0.807, respectively. This finding further shows that climate-related factors have a significant influence on borrowers' repayment capacity in the agricultural context. Furthermore, the S+C model demonstrates a strong correlation (0.936) with the S+C+T model. This result suggests that climate data continues to drive default risk signals, even in the presence of additional information from text data. This finding is consistent with our earlier prediction performance results as outlined in Section 6.2.

To better understand the role of climate risk factors in predicting credit default, we employ SHAP to analyse feature importance and temporal dependencies (Korangi et al., 2023). Following Stevenson et al. (2021) and Wu et al. (2025), we focus on uncertain cases in the test set, identified as those where the structured-only model yields predicted probabilities in the middle range (30th to 70th percentiles), while an improvement in the predicted probabilities is observed when using the combined structured and climate (S+C) model. For this subset, we review 491 cases demonstrating the greatest improvement under the S+C model. Figure 4 presents the average contribution of each climate risk factor to the predicted default probabilities. As shown in Figure 4, the water-logging by rain risk exhibits the most significant relative contribution. This pattern is probably due to the fact that excessive rainfall and flooding events can severely disrupt agricultural productivity by damaging crops, reducing yields, and diminishing the value of agricultural collateral, ultimately influencing the borrowers' repayment capabilities.

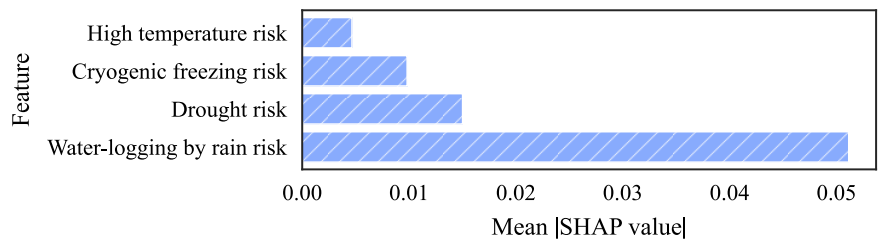

**Figure 4** Average contribution of each climate risk factor

To further explore the temporal effects of climate risk factors, we randomly selected two time periods to visualise the SHAP value distributions. Figure 5 presents the SHAP value distributions for climate risk factors measured at t-1, which corresponds to the month immediately preceding the loan start date, and t-8, which corresponds to eight months prior to loan origination. The results in Figures 5(a) and 5(b) reveal a positive correlation between heightened water-logging by rain risk levels and increased default likelihood. Drought risk appears as the second most influential climate factor; nevertheless, its relationship with default probability exhibits a generally opposite pattern to that of the water-logging by rain risk. Specifically, Figures 5(a) and 5(b) show that lower drought risk levels can sometimes correspond to higher default probabilities. One plausible explanation is that reduced drought stress may increase vulnerability to flooding. For example, higher soil moisture levels can exacerbate water-logging during periods of excessive rainfall. Additionally, regions with relatively low drought risk may still face indirect vulnerabilities, such as inefficient water management or over-irrigation, which could result in crop damage, soil degradation, or compromised agricultural productivity. These effects may increase repayment difficulties despite lower observed drought risk, which highlights the importance of considering multiple climate dimensions simultaneously in credit risk assessments. Although high-temperature and cryogenic freezing risks exhibit smaller contributions than water-logging and drought risks, they still influence default probabilities. High-temperature risk is associated with crop stress, yield reductions, and heat-related damage to agricultural infrastructure, which can indirectly affect borrowers' financial stability. Similarly, cryogenic freezing risk captures the potential for frost damage, particularly in regions susceptible to unexpected cold events.

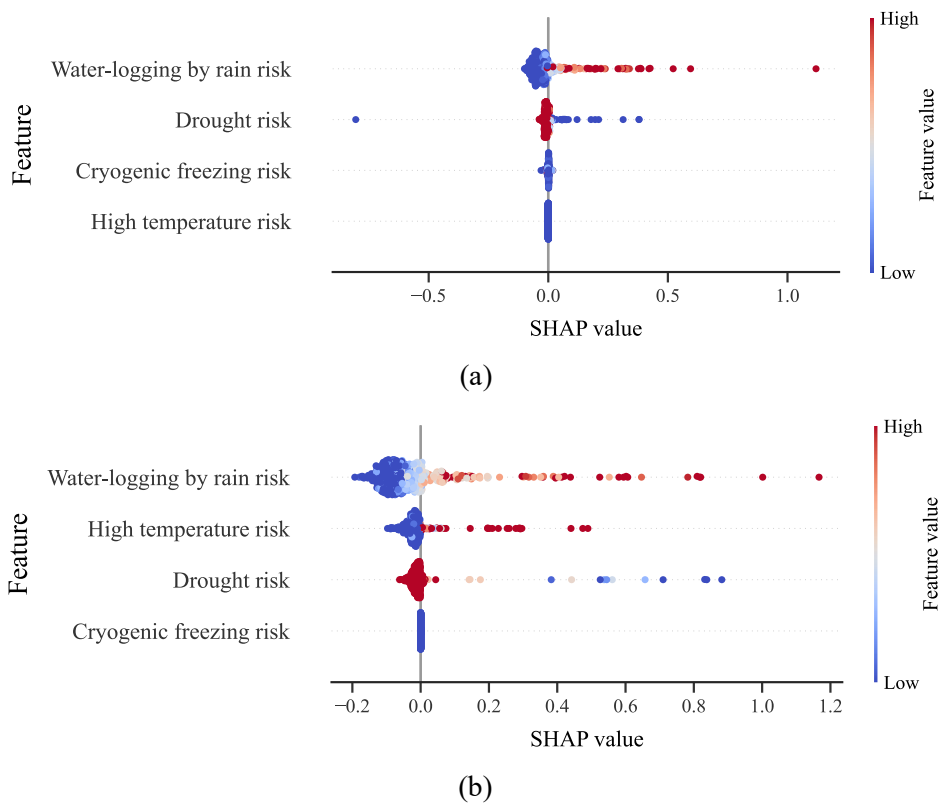

**Figure 5** SHAP value distribution for climate risk factors across periods: (a) t-1, (b) t-8

We further investigate the individual contribution of each climate factor to default prediction by integrating them into the models alongside the structured data. Table 5 summarises the discrimination performance across the different data modalities. The results show that the model integrating the structured data with water-logging by rain risk achieves the highest average AUC (0.667), KS (0.381), and H-measure (0.196) among all tested models. This finding is also consistent with our earlier SHAP analysis, where water-logging by rain risk exhibited the highest mean SHAP value. Moreover, we can observe that the models incorporating high-temperature and drought risks demonstrate relatively strong predictive performance, with results comparable to those of the structured-only model. In contrast, the model combining the structured data with cryogenic freezing risk produces relatively lower results across all metrics. These findings emphasise the varying degree of influence of each climate risk factor on default prediction and highlight the particular importance of water-logging by rain risk in improving model performance in the agricultural setting.

**Table 5** Model performance across different data modalities (structured+individual climate factor)

| Modality | AUC | KS | H-measure |
| --- | --- | --- | --- |
| S+water-logging by rain risk | **0.667 (0.665, 0.669)** | **0.381 (0.379, 0.384)** | **0.196 (0.194, 0.199)** |
| S+high-temperature risk | 0.617 (0.615, 0.619) | 0.299 (0.297, 0.301) | 0.176 (0.174, 0.178) |
| S+drought risk | 0.608 (0.606, 0.609) | 0.296 (0.294, 0.298) | 0.161 (0.159, 0.163) |
| S+cryogenic freezing risk | 0.581 (0.579, 0.582) | 0.260 (0.258, 0.262) | 0.116 (0.115, 0.118) |

**Notes**: This table presents the discrimination performance, as shown by the AUC, KS, and H-measure (mean and its 95% confidence interval), in terms of predicting the default risk of borrowers based on four data modalities: S+water-logging by rain risk, S+high-temperature risk, S+drought risk, and S+cryogenic freezing risk.

To evaluate the impact of adding each climate risk factor to the structured data and the shifts in relative modal importance, we also review the correlations between the different sets of model outputs in Table 6. The results demonstrate that most of the combined models involving the different climate risk factors exhibit high correlations with the structured-only model, as we might expect given the richness of the structured features. However, the model combining the structured data with water-logging by rain risk displays a moderate correlation of 0.426 with the structured-only model. This result suggests that the inclusion of water-logging by rain risk introduces unique information that differs from the information obtained from the structured data, which highlights the distinctive influence of water-logging by rain risk in capturing default prediction nuances in the agricultural setting.

**Table 6** Spearman's rank correlations between predicted default probabilities (structured+individual climate factor, test set)

| Modality | S | S+water-logging by rain risk | S+high-temperature risk | S+drought risk | S+cryogenic freezing risk |
| --- | --- | --- | --- | --- | --- |
| S | 1.000 | | | | |
| S+water-logging by rain risk | 0.426 | 1.000 | | | |
| S+high-temperature risk | 0.955 | 0.461 | 1.000 | | |
| S+drought risk | 0.985 | 0.395 | 0.936 | 1.000 | |
| S+cryogenic freezing risk | 0.775 | 0.523 | 0.793 | 0.737 | 1.000 |

## 7. Conclusions

The increasing availability of multimodal data offers promising avenues for making breakthroughs in credit default prediction, particularly for mSEs that often suffer from limited or incomplete financial records. Among the growing range of alternative data sources, climate data has attracted increasing attention because of its direct and indirect effects on borrowers' operating conditions and repayment capacity, especially in climate-sensitive sectors such as agriculture (Castro & Garcia, 2014; Pelka et al., 2015; Römer & Musshoff, 2018). At the same time, textual data offer complementary insights by capturing qualitative and context-specific information that is difficult to quantify using structured financial variables alone (Stevenson et al., 2021; Katsafados et al., 2024; Wu et al., 2025). Despite this potential, existing studies have largely examined these data sources in isolation, and relatively little attention has been paid to how different modalities can be systematically integrated within a unified modelling framework.

With the technology development, recent advances in deep learning provide powerful tools for extracting and combining information from heterogeneous data sources. In this study, we propose a deep learning-based multimodal framework that jointly integrates structured financial data, climate panel data, and unstructured textual narratives. In our empirical experiments, we use three deep learning models, LSTM, GRU, and transformer-based models, to validate the effectiveness of the proposed multimodal architecture in enhancing credit default prediction. Our findings first reveal that models relying solely on climate or text data significantly outperform those based on structured data. These results highlight the predictive value of alternative data sources, particularly climate data, for enhancing credit default prediction performance within agricultural settings. Furthermore, the integration of multiple data modalities significantly improves prediction accuracy beyond what can be achieved using any single modality, with transformer-based models achieving superior results. These findings suggest that different data sources convey complementary information about default risk and that combining diverse data modalities is an effective strategy for more accurately predicting credit risk. Moreover, our study provides robust evidence that highlights the critical role climate risk factors play in shaping borrower repayment behaviours. By employing correlation analysis and SHAP methods for interpretability, we identify water-logging by rain risk as the most significant climate risk factor. Excessive rainfall can disrupt agricultural productivity and diminishes collateral value, thereby weakening borrowers' repayment capacity.

Our research not only contributes to the growing literature on the application of multimodal learning in the business and IS domains but also has far-reaching managerial implications. First, for financial institutions, the proposed multimodal framework enables a more comprehensive assessment of credit risk by integrating structured credit variables with climate panel measures and loan-officer narratives. This richer information set can help improve both borrower screening and post-origination monitoring, thereby reducing default risk and enhancing portfolio stability (Wu et al., 2025). In agricultural lending in particular, lenders can exploit the incremental value of multimodal data to refine approval thresholds, adjust risk-based pricing, and strengthen early warning systems. The strong predictive role of climate factors also suggests that lenders can implement climate-aware risk segmentation and adapt covenants, collateral requirements, or insurance arrangements in regions exposed to different climate risks.

Furthermore, for regulators and policymakers, our findings suggest that physical climate exposure can be a first-order driver of default risk in certain portfolios. This evidence supports the development of supervisory guidance and lending regulations that explicitly account for environmental stressors (Campiglio et al., 2018; Dikau & Volz, 2021). Encouraging financial institutions to systematically incorporate climate factors into credit assessments can strengthen financial system resilience, promote climate-aware risk management, and support sustainable lending practices in climate-sensitive sectors. Moreover, by leveraging alternative data sources, lenders can better assess the risk profiles of firms with limited financial histories, which facilitates fairer access to credit, supports tailored financing programmes, and enhances the resilience of local economies.

Finally, for IS stakeholders and designers, this study shows how multimodal risk modelling can be implemented as a modular decision-support artefact (Phillips-Wren, 2013; Watson, 2017). The modality-specific encoding and fusion architecture supports maintainability and extensibility, which allows institutions to add, update, or remove data channels without rebuilding the entire scoring pipeline. As alternative data sources, such as text, images, audio, and video, become increasingly available, our framework provides a practical blueprint for integrating diverse information streams into multimodal systems to support operational and strategic decision-making.

While this study offers valuable insights into the integration of multimodal data for credit risk prediction, several limitations highlight opportunities for future research. First, an interesting avenue for further research would be to extend the multimodal learning architecture proposed in our paper by incorporating additional data modalities, such as real-time satellite imagery, borrowers' network data, and social media activity. Although previous research has highlighted the value of such unstructured data, little work has been done to integrate these sources alongside structured data for enhanced credit default prediction. Second, our analysis is limited to four climate-related risk factors. Expanding this scope to include additional natural hazards, such as earthquakes, landslides, and hurricanes, could facilitate a more comprehensive understanding of the interplay between environmental risks and credit defaults. Third, while our framework performs well on a dataset of agricultural mSEs, its scalability and applicability to larger datasets or other loan types, remain unexplored. Future research should evaluate the robustness of these models across diverse contexts, including varying loan durations and macroeconomic conditions.

**Declarations**

Ethics approval and consent to participate: None to declare.

Consent for publication: All authors of the manuscript give consent to publish this article.

Availability of data and materials: The data cannot be shared due to confidentiality agreements. The code can be shared upon request for academic and research purposes only.

Competing interests: None to declare.

Funding: Not applicable.

Authors' contributions: Zongxiao Wu – conceptualisation, methodology, software, validation, formal analysis, investigation, resources, writing - original draft, writing - review & editing, visualisation. Ran Liu – methodology, software, formal analysis, data curation, writing - original draft, visualisation. Jiang Dai – writing - original draft, writing - review & editing, project administration. Dan Luo – writing - review & editing, project administration.

Acknowledgement: None to declare.

# Supplementary materials

**Appendix A. Literature review**

**Table A1** Literature on the integration of climate data into credit default prediction

| Author | Data type | Climate category | Model |
|---|---|---|---|
| Collier et al. (2011) | Microfinance | Floods | Time-series model |
| David (2011) | Macroeconomic | Climatic events (floods, droughts, extreme temperatures, and hurricanes), geological events (earthquakes, landslides, volcano eruptions, and tidal waves), and human disasters (famines and epidemics) | The generalised method of moments |
| Berg and Schrader (2012) | Corporate loan | Volcanic eruptions | LPM |
| Castro and Garcia (2014) | Agricultural loan | Temperatures and precipitation | Generalised linear model |
| Klomp (2014) | Bank distress | Hydrological disasters (floods and wet mass movements), meteorological disasters (storms and hurricanes), geophysical disasters (earthquakes, tsunamis, and volcanic eruptions), extreme temperatures, droughts, and wildfires | Dynamic panel model |
| Pelka et al. (2015) | Agricultural loan | Precipitation | LPM, Sequential logit model |
| Römer and Mußhoff (2018) | Agricultural loan | Precipitation | Logistic regression |
| Brei et al. (2019) | Bank distress | Hurricanes | Panel regression model |
| Issler et al. (2020) | Mortgage loan | Wildfires | Game-theoretic model |
| Kaur Brar et al. (2021) | Agricultural loan | Temperatures | Portfolio optimisation model |
| Ouazad and Kahn (2021) | Mortgage loan | Floods | LRM |
| Choudhary and Jain (2022) | Consumer loan | Floods | LRM |
| Breeden (2023) | Macroeconomic | Droughts | Macroeconomic stress model |
| Gao et al. (2023) | Agricultural loan | Severe weather (hail, thunderstorms, high winds, droughts, flash floods, winter storms, heat, heavy snows, winter weather, tornadoes, floods, strong winds, and marine thunderstorms) | Neural network, extreme gradient boosting, random forest |
| Abedifar et al. (2024) | Agricultural loan | Floods | Difference-in-difference regression model |
| Calabrese et al. (2024) | Mortgage loan | Heavy rains and tropical cyclones | Additive Cox proportional hazard model |
| Lane (2024) | Agricultural loan | Floods | LRM |
| Chen et al. (2025) | Bank distress | Temperatures, precipitation, and carbon emissions | Spatial analysis model |

**Abbreviations**: LPM - linear probability model, LRM - linear regression model

# Appendix B. Overview of structured features

**Table B1** Definitions and statistical information for structured features

| Feature | Definition | Summary statistics | | | | |
|---|---|---|---|---|---|---|
| *Continuous* | | *Obs* | *Mean* | *SD* | *Min* | *Max* |
| Age | The age of the business owner | 4,172 | 40.83 | 8.43 | 20.00 | 64.00 |
| Annual expense ♣ | Total annual expense (¥) | 4,172 | 18,306.00 | 38,162.44 | 5,456.70 | 330,000.00 |
| Annual revenue | Total annual revenue (¥) | 4,172 | 179,133.58 | 344,940.88 | 32,400.00 | 2,364,000.00 |
| Bedrooms ♣ | Number of bedrooms in the borrower's house | 4,172 | 4.84 | 2.59 | 1.00 | 16.00 |
| Family members ♣ | Number of family members of the borrower | 4,172 | 3.57 | 0.93 | 1.00 | 9.00 |
| Family workforce ♣ | Number of members of the workforce available in the borrower's family | 4,172 | 2.48 | 0.77 | 1.00 | 8.00 |
| Floors ♣ | Number of floors in borrower's house | 4,172 | 1.72 | 0.82 | 1.00 | 5.00 |
| House area | Borrower's residential land area ($m^2$) | 4,172 | 174.85 | 132.93 | 50.00 | 1,064.70 |
| Loan amount ♣ | Loan amount (¥) | 4,172 | 39,368.12 | 15,665.30 | 1,000.00 | 100,000.00 |
| Loan term ♣ | Loan term (months) | 4,172 | 3.85 | 0.69 | 1.00 | 9.00 |
| Monthly revenue | Average monthly revenue (¥) | 4,172 | 15,804.74 | 12,303.22 | 7,135.00 | 91,999.92 |
| Rate-of-income | The ratio of monthly repayment to income (%) | 4,172 | 12.65 | 33.27 | 2.00 | 408.90 |
| Rest amount | Remaining loan amount outstanding (¥) | 4,172 | 482.46 | 4,504.41 | 0.00 | 80,000.00 |
| Rest interest | Remaining loan interest outstanding (¥) | 4,172 | 24.03 | 282.34 | 0.00 | 6,770.93 |
| *Categorical* | | | *Number of categories* | | *Values* | |
| Business type ♣ | Type of borrower's business | 4,172 | 3 | | {1, 2, 3} | |
| Credit rating ♣ | Borrower's credit rating | 4,172 | 6 | | {1, 2, 3, 4, 5, Other} | |
| Customer type | Whether the borrower is a new or existing customer of the bank | 4,172 | 2 | | {New_cust, Old_cust} | |
| Degree ♣ | Borrower's academic degree | 4,172 | 4 | | {0, 4, 5, 9} | |
| Education ♣ | Borrower's educational background | 4,172 | 10 | | {10, 20, 30, 40, 50, 60, 70, 80, 90, 99} | |
| Ethnic group | Borrower's ethnic group | 4,172 | 10 | | {A, B, C, D, E, F, G, H, I, J} | |
| Homeownership ♣ | Borrower's homeownership | 4,172 | 6 | | {1, 2, 3, 4, 5, Other} | |

| | | | | |
|---|---|---|---|---|
| House type ♣ | House type | 4,172 | 2 | {Flat, House} |
| Housekeeping ♣ | Housekeeping condition | 4,172 | 3 | {Bad, Moderate, Good} |
| Job position ♣ | Borrower's job position | 4,172 | 6 | {1, 2, 3, 4, 5, Other} |
| Job title | Borrower's job title | 4,172 | 6 | {1, 2, 3, 4, 5, Other} |
| License type ♣ | Type of business license | 4,172 | 11 | {A, C, E, F, H, O, P, Q, S, Z, Other} |
| Marital relationship ♣ | Whether the borrower's marital relationship is harmonious | 4,172 | 4 | {Bad, Moderate, Good, Very good} |
| Marital status ♣ | Borrower's marital status | 4,172 | 7 | {10, 21, 22, 23, 30, 40, 90} |
| Occupation ♣ | Borrower's occupation | 4,172 | 10 | {0, 1, 3, 4, 5, 8, 9, X, Y, Z} |
| Postcode | The postcode of the business location | 4,172 | 12 | {153200, 065300, 164100, 325700, …}* |
| Repay type ♣ | Type of loan repayment | 4,172 | 3 | {A, B, C} |
| Verified ID ♣ | Borrower's citizenship category | 4,172 | 7 | {1, 2, 3, 4, 5, 6, Other} |

**Notes**: We conducted a data desensitisation process on the original structured data to protect private and business-sensitive information. * indicates that some infrequent categories are omitted in the interest of space. ♣ indicates the final features selected for our models.

## Appendix C. *SPI* calculation

The standardised precipitation index (*SPI*) is calculated by the following steps:

First, we assume that the precipitation $x$ during a given time period conforms to a $\Gamma$ distribution:

$$f(x) = \frac{1}{\beta^\gamma \Gamma(\gamma)} x^{\gamma-1} e^{-\frac{x}{\beta}}, x > 0 \tag{C1}$$

where $\gamma = \frac{1+\sqrt{1+\frac{4A}{3}}}{4A}$ $(A = lg\bar{x} - \frac{1}{n} n \sum_{i=1}^{n} lgx_i)$, $\beta = \frac{\bar{x}}{\gamma}$.

For the annual precipitation $x_0$ in a certain year, the probability of the random variable $x$ can be calculated using Equation (C2), where $m$ is the number of samples with zero precipitation, and $n$ represents the total number of samples.

$$F(x) = \begin{cases} \int_0^{x_0} f(x)dx, & x < x_0 \\ \frac{m}{n}, & x = 0 \end{cases} \tag{C2}$$

By substituting the probability values obtained from Equation (C2) into the $\Gamma$ distribution, we obtain

$$F(x < x_0) = \frac{1}{\sqrt{2\pi}} \int_0^{x_0} e^{-\frac{z^2}{2}} dx \tag{C3}$$

Finally, we calculate the *SPI* as follows:

$$SPI = S\left[t - \frac{(c_2 t + c_1)t + c_0}{((d_3 t + d_2)t + d_1)t + 1}\right] \tag{C4}$$

where $c_0 = 2.515517$, $c_1 = 0.802853$, $c_2 = 0.010328$, $d_1 = 1.432788$, $d_2 = 0.189269$, $d_3 = 0.001308$, $t = \sqrt{ln\frac{1}{F^2}}$ and $S = 1$ when $F > 0.5$ and $S = -1$ when $F \leq 0.5$.

**Appendix D. Robustness checks**

In this appendix, we additionally report results from the main experiments using logistic regression (LR) and extreme gradient boosting (XGB) in Table D1, and we implement a 5-fold cross-validation strategy in Table D2. Overall, these results are consistent with our main findings. In particular, models that integrate multiple data modalities consistently outperform unimodal specifications, which confirms the added predictive value of multimodal learning. Among the alternative data sources, the inclusion of climate risk factors yields the most pronounced performance improvements. These findings suggest that our results are robust to alternative classifiers and validation strategies.

**Table D1** Model performance across different data modalities (LR and XGB)

| Model | Modality | AUC | KS | H-measure |
|---|---|---|---|---|
| LR | Structured | 0.606 (0.604, 0.607) | 0.290 (0.288, 0.292) | 0.157 (0.155, 0.159) |
|  | Climate | 0.728 (0.726, 0.729) | 0.438 (0.435, 0.440) | 0.250 (0.248, 0.252) |
|  | Text | 0.628 (0.626, 0.629) | 0.307 (0.305, 0.309) | 0.129 (0.127, 0.130) |
|  | Structured+Climate | 0.747 (0.746, 0.748) | **0.481 (0.479, 0.483)** | 0.268 (0.266, 0.270) |
|  | Structured+Text | 0.609 (0.607, 0.611) | 0.283 (0.281, 0.284) | 0.161 (0.160, 0.163) |
|  | Climate+Text | 0.731 (0.729, 0.732) | 0.456 (0.454, 0.459) | 0.254 (0.252, 0.256) |
|  | Structured+Climate+Text | **0.748 (0.747, 0.750)** | 0.480 (0.478, 0.482) | **0.269 (0.267, 0.271)** |
| XGB | Structured | 0.640 (0.638, 0.642) | 0.299 (0.297, 0.302) | 0.147 (0.145, 0.149) |
|  | Climate | 0.713 (0.712, 0.715) | 0.404 (0.401, 0.407) | 0.225 (0.223, 0.227) |
|  | Text | 0.629 (0.627, 0.630) | 0.316 (0.314, 0.319) | 0.156 (0.155, 0.158) |
|  | Structured+Climate | **0.743 (0.741, 0.744)** | **0.462 (0.460, 0.463)** | 0.258 (0.257, 0.259) |
|  | Structured+Text | 0.741 (0.739, 0.742) | 0.460 (0.458, 0.462) | 0.252 (0.250, 0.253) |
|  | Climate+Text | 0.638 (0.636, 0.640) | 0.314 (0.312, 0.316) | 0.188 (0.186, 0.190) |
|  | Structured+Climate+Text | 0.647 (0.645, 0.649) | 0.359 (0.356, 0.362) | **0.261 (0.258, 0.263)** |

**Notes**: This table presents the discrimination performance, calculated using the AUC, KS, and H-measure (mean and its 95% confidence interval), of two model architectures: LR and XGB, used for predicting the default risk of borrowers.

**Table D2** Model performance across different data modalities (5-fold cross validation)

| Model | Modality | AUC | KS | H-measure |
|---|---|---|---|---|
| LSTM | Structured | 0.643 (0.640, 0.646) | 0.374 (0.371, 0.378) | 0.217 (0.214, 0.220) |
|  | Climate | 0.652 (0.650, 0.655) | 0.380 (0.376, 0.383) | 0.245 (0.242, 0.247) |
|  | Text | 0.588 (0.585, 0.591) | 0.315 (0.311, 0.318) | 0.177 (0.173, 0.180) |
|  | Structured+Climate | **0.748 (0.746, 0.750)** | **0.489 (0.486, 0.492)** | **0.367 (0.364, 0.371)** |
|  | Structured+Text | 0.691 (0.688, 0.693) | 0.415 (0.412, 0.418) | 0.255 (0.252, 0.258) |
|  | Climate+Text | 0.583 (0.581, 0.586) | 0.301 (0.298, 0.304) | 0.189 (0.186, 0.192) |
|  | Structured+Climate+Text | 0.715 (0.713, 0.718) | 0.464 (0.460, 0.467) | 0.297 (0.293, 0.300) |
| GRU | Structured | 0.643 (0.640, 0.646) | 0.374 (0.371, 0.378) | 0.217 (0.214, 0.220) |
|  | Climate | 0.670 (0.667, 0.673) | 0.404 (0.400, 0.408) | 0.269 (0.266, 0.271) |
|  | Text | 0.620 (0.617, 0.623) | 0.340 (0.337, 0.344) | 0.230 (0.227, 0.234) |
|  | Structured+Climate | 0.691 (0.689, 0.693) | 0.415 (0.412, 0.418) | 0.266 (0.263, 0.269) |
|  | Structured+Text | 0.705 (0.703, 0.708) | **0.463 (0.460, 0.467)** | 0.303 (0.299, 0.306) |
|  | Climate+Text | 0.676 (0.673, 0.679) | 0.421 (0.417, 0.425) | 0.295 (0.291, 0.299) |
|  | Structured+Climate+Text | **0.725 (0.722, 0.728)** | 0.462 (0.458, 0.466) | **0.342 (0.338, 0.346)** |
| Transformer | Structured | 0.643 (0.640, 0.646) | 0.374 (0.371, 0.378) | 0.217 (0.214, 0.220) |
|  | Climate | 0.679 (0.677, 0.682) | 0.407 (0.404, 0.411) | 0.269 (0.266, 0.272) |

| | | | |
|---|---|---|---|
| Text | 0.657 (0.654, 0.661) | 0.413 (0.409, 0.416) | 0.293 (0.289, 0.296) |
| Structured+Climate | 0.746 (0.744, 0.749) | 0.503 (0.500, 0.506) | 0.352 (0.349, 0.355) |
| Structured+Text | 0.757 (0.754, 0.760) | 0.513 (0.511, 0.514) | 0.361 (0.358, 0.364) |
| Climate+Text | 0.728 (0.726, 0.730) | 0.475 (0.472, 0.478) | 0.330 (0.327, 0.333) |
| Structured+Climate+Text | **0.760 (0.758, 0.762)** | **0.519 (0.516, 0.523)** | **0.369 (0.366, 0.372)** |

**Notes**: This table presents the discrimination performance, calculated using the AUC, KS, and H-measure (mean and its 95% confidence interval), of three model architectures: LSTM, GRU, and transformer, used for predicting the default risk of borrowers.